\documentclass[11pt]{article}

\usepackage{amsmath}
\usepackage{graphicx}
\usepackage{setspace}
\usepackage{float}
\usepackage{color}

\begin{document}

\title{Well-temperate phage: 
optimal bet-hedging 
against local environmental collapses}

\author{Sergei Maslov$^{1}$ \\ Kim Sneppen$^{2}$
\\\\
\small $^{1}$ Biological, Environmental and Climate Sciences Department,  \\
\small Brookhaven National Laboratory, \\
\small  Upton, NY 11973, USA\\\\
\small $^{2}$ Center for Models of Life, Niels Bohr Institute, \\
\small University of Copenhagen, 2100 Copenhagen, Denmark}
\date{\small \today}
\maketitle

\begin{center}
Correspondence to: \\
Kim Sneppen: $ksneppen@gmail.com$\\
Sergei Maslov: $ssmaslov@gmail.com$
\end{center}

\begin{doublespace}
\begin{abstract}
Upon infection of their bacterial hosts temperate
phages must chose between lysogenic and lytic developmental strategies.
Here we apply the game-theoretic bet-hedging strategy 
introduced by Kelly to derive 
the optimal lysogenic fraction 
of the total population of phages as a function of frequency and intensity of 
environmental downturns affecting the lytic subpopulation.  
``Well-temperate'' phage from our title is
characterized by the best long-term population 
growth rate. 
We show that it is realized when the
lysogenization frequency is 
approximately equal to the probability of 
lytic population collapse.
We further predict the existence of sharp boundaries 
in system's environmental, ecological, and biophysical 
parameters separating the regions 
where this temperate strategy is optimal from 
those dominated by purely virulent 
or dormant (purely lysogenic) strategies. We show that 
the virulent strategy works best for phages 
with large diversity of hosts, and
access to multiple independent environments reachable by diffusion.
Conversely, progressively more temperate or even dormant
strategies are favored in the environments, that are subject to frequent and 
severe temporal downturns.
\end{abstract}

\newpage

\section*{Author Summary}
Coexistence between bacteria and 
bacteriophages gives rise to 
a variety of strategies each organism uses 
to exploit weaknesses of another.
In particular, while virulent phages always 
kill their host by lysing or breaking it open,
temperate phages can choose between such lytic 
strategy and entering dormant or 
lysogenic state in which their
host is maintained nearly intact. 
While many biomolecular mechanisms 
regulating lytic-to-lysogenic transition are well 
understood, there is no clear agreement on environmental 
properties favoring virulent over temperate phage strategy.
Also, currently there is no general theoretical framework 
relating the frequency of lysogeny for temperate phages 
to contemporary environmental conditions.
Here we address both questions using 
the classical bet-hedging strategy first introduced by Kelly. 
Our analysis shows that the best long-term growth rate of the phage
population is achieved by Kelly-optimal or "well-tempered" strategy 
from our title with lysogenization frequency approximately equal to 
the probability of lytic subpopulation collapse.
The fitness advantage of this optimized strategy over purely virulent 
one can be expressed in terms of  a relative information entropy.
We predict sharp transitions between optimal virulent, 
temperate, and dormant (purely lysogenic) strategies 
as a function of environmental, ecological, and biophysical 
parameters of the system. 
In particular,
purely virulent strategy is shown to work best for phages 
with large diversity of hosts, and
access to multiple independent environments reachable by diffusion.
Conversely, temperate or dormant strategies are favored in 
environments subject to frequent and severe temporal downturns.
\newpage

\section*{Introduction}

Bacteria and their main predators, bacteriophages \cite{campbell1960,valera},  
are the most abundant and dynamic part of the biosphere 
\cite{bergh1989,whitman1998}. 
Phages lead a risky lifestyle 
\cite{campbell1960,jessup2008,weitz2005}
and as a consequence local populations of individual phage species 
routinely experience extreme fluctuations caused by changes 
in availability of susceptible hosts \cite{wiggins1985,buckling2002}.
In real ecosystems these fluctuations may be caused 
e.g. by depletion of nutrients for hosts,
development of host resistance, and 
interference from competing phages \cite{haerter2014}
or other bacterial predators. 

Phages deal with these challenges using a variety 
of strategies \cite{campbell1960,jessup2008}.
One common strategy adopted by virulent phages 
is to always kill and lyse their host resulting in release
of 100-3000 progeny phages \cite{paepe2006}.
Sustainability of this strategy is critically
dependent on phages' ability to reach susceptible hosts 
\cite{campbell1960,levin1977,kerr2002,weitz2005,wiggins1985,paepe2006}
because free phage particles have a finite lifetime \cite{paepe2006}. 
In contrast, temperate phages following the infection
of a bacterium can opt for a transition to the lysogenic state 
where the future fate of the incorporated prophage is aligned 
with that of its bacterial hosts.
Ref. \cite{stewart1984} suggested that
the temperate strategy may persist because it allows
phages to survive extended periods 
when host density is below the level needed to sustain
the propagation of the lytic population (pure virulent strategy). 

Campbell \cite{campbell1960} considered the sustainability of 
pure virulent strategy, noting that it 
strongly depends on the growth rate of the phages' bacterial host 
relative to that of other competing bacteria in the same local environment.
Thus, any changes in the environment affecting the relative 
ranking of bacteria by their growth rate will likely impact 
the local success of lytic phages infecting these bacteria
\cite{haerter2014}. 

The lysogenic state of temperate phages allows them to 
weather out severe downturns in environmental conditions 
\cite{stewart1984,avlund2009}. However, it comes at the cost of some 
reduction in the growth rate of the lytic subpopulation. 
It is hence plausible that 
temperate phages should try to optimize the ratio of their populations in lytic and
lysogenic states in order to achieve the maximal long-term growth rate.
To quantify this process we consider a local phage population 
living in a fluctuating environment characterized by sudden 
unpredictable downturns, a scenario inspired
by the classical paper by Kelly \cite{kelly1956} on application 
of information theory to gambling. Like gamblers \cite{kelly1956} 
or financial investors \cite{maslov1998} phages must decide
what part of their population ``capital'' to allocate to a ``risky'' 
lytic state which has the potential for rapid growth 
but is also subject to a non-negligible risk of sudden collapse. 
The rest of the population ``capital'' will be allocated to 
the relatively ``safe'' lysogenic state. Here we aim to quantify 
the parameters of the long-term optimal or ``well-temperate'' phage 
strategy and compare its outcome to a purely virulent strategy.

Our calculations are inspired by the original work by Kelly \cite{kelly1956}, 
its applications to finance  \cite{maslov1998, marsili1998}
and evolution \cite{bergstrom2004}. Similar bet-hedging
approaches have been explored to germination of
seeds from annual plants \cite{cohen1966,seger87,bulmer1984}.
As is common for bet-hedging in population biology, 
our analysis goes beyond optimization of the outcome 
of individual infection on a short-time scale,
and instead considers the long-term (logarithmic) growth rate
\cite{cohen1966,metz1992}.

\section*{Results}

\subsection*{ The Kelly-optimal frequency of lysogeny}

Consider a local phage population that grows or declines in fluctuating environmental conditions. 
The environment is assumed to randomly switch between ``good" conditions favoring multiplicative growth of the 
lytic subpopulation 
and ``bad'' conditions during which local 
lytic subpopulation completely (or partially
as will be investigated later in this study) dies out. Rapid growth during good conditions is quantified 
by the amplification factor $\Omega >1$. We assume bad conditions to be 
transient events of indefinite duration that occur with the probability $p \ll 1$. 
During good environmental conditions one time step of our model roughly corresponds to one phage generation 
which in turn makes $\Omega$ bounded from above by phage's burst size.
Conversely, during bad environmental 
conditions we 
count the entire duration of this event as a single time step.

Temperate phages in our model have no 
predictive knowledge of when their environment is about to turn bad. 
However, they are free to choose what fraction $x$ of their population will be 
kept in the lysogenic state at every time step.
Notice that because only during new infections by phages from the lytic subpopulation 
they are free to chose between lytic and lysogenic states, our model is based on the 
assumption that the majority of phages are of this type. This assumption 
is justified in case of rapid exponential growth when the 
lysogenization frequency during the last ``good'' time-step  
approximately determines the lysogenic fraction $x$ 
for the entire population.  In the simplest scenario considered in this 
chapter we also assume that the local lysogenic subpopulation is fully protected
from the extreme changes in the local environment and does not 
change with time. Later on in this study we will relax this assumption
 and allow the lysogenic population to be characterized by a time-independent 
growth (or decline) rate.  In fact the growth rate of the lytic subpopulation 
is always defined {\it relative} to that of the lysogenic subpopulation 
much in the same way as in financial markets risky asset returns 
are always compared to interest rates paid by banks.
The expected value of the (logarithmic)
growth rate $\Lambda$ of the entire local phage population in our model is given by
\begin{equation}
\Lambda(x)=(1-p) \log \left[(1-x)\Omega +x\right]+p \log x \label{eq1}
\end{equation} 
The first term is the logarithmic growth rate under good conditions when the lytic
fraction $1-x$ of the total population is multiplied by $\Omega$, 
while the lysogenic subpopulation $x$ remains unchanged.
The second term is the logarithmic growth rate under bad conditions when only phages in the
lysogenic state survive. Later on we will relax the requirement that the entire lytic population 
has to completely die off during bad times. 
The growth rate considered above weights the {\it logarithms} of multiplicative 
growth factors of the entire phage population under two conditions 
with their respective probabilities of occurrence. Maximization of $\Lambda$ with respect to 
$x$ secures the long-term optimal growth rate \cite{kelly1956}. This should 
not be confused with optimization of the expected (average) population growth after just
one or a small number of growth cycles. Such short-term optimization would always favor purely lytic 
strategy with $x=0$ provided that $(1-p)\Omega>1 $. The last condition is almost always
fulfilled since during good times $\Omega$ approaches its upper bound given by the 
average burst size which is substantially larger than one offspring per phage. On the other hand, 
following the lytic strategy for a long time would almost certainly bring 
the phage population to the total collapse, which will happen during the very 
first bad time interval.

\begin{figure}[H]
\includegraphics[trim = 0mm 0mm 0mm 0mm, clip,angle=270,width=\columnwidth]{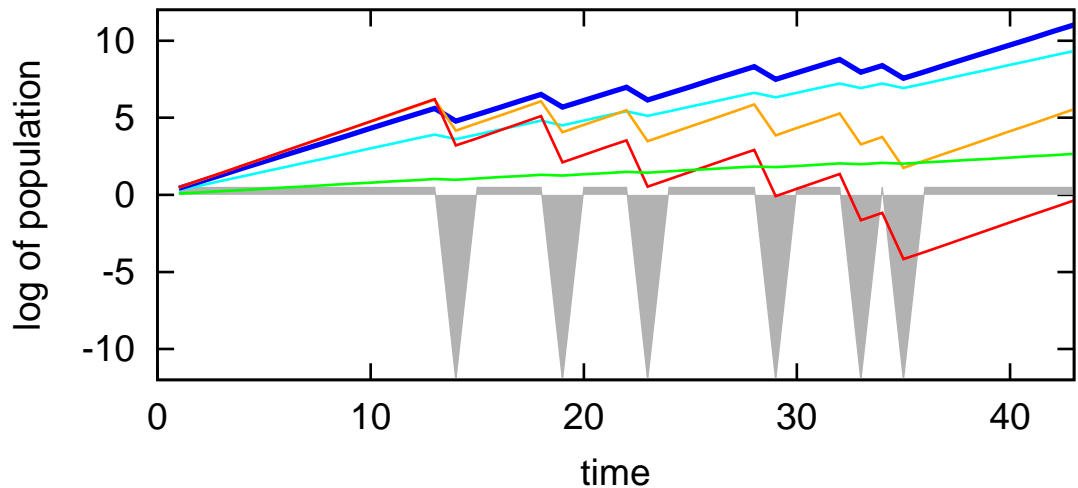}
\caption{\small\sl Phage population dynamics 
when exposed to long periods of exponential growth with 
$\Omega=3$ interrupted by occasional bad conditions
where lytic subpopulation drops nearly to zero.
Note the logarithmic scale base 10 on the y-axis.
Bad conditions of severity $\omega=10^{-12}$ happening 
with probability $p=0.1$ are marked
with downward-facing grey triangles.
The blue curve is the growth of the phage population following the
Kelly-optimal strategy with lysogenic fraction $x^* =p \cdot \Omega/(\Omega-1)=0.15$,
whereas orange and red curves show suboptimal strategies 
with $x=0.01$ and $x=0.001$ correspondingly. 
Conversely, cyan and green curves simulate phage 
population dynamics with higher-than-optimal lysogenization 
frequencies of respectively $x=0.5$ and $x=0.9$. 
In the long run, phages following the Kelly-optimal
strategy outperform their competitors. 
Note that the growth rate of the entire phage 
population is measured relative to that of the lysogenic 
subpopulation. If the latter is negative one can have 
a steady state Kelly-optimal solution (as opposed to the unlimited 
exponential growth shown here).
}
\label{figure1}
\end{figure}

In contrast to its short-term counterpart, the long-term logarithmic growth rate $\Lambda(x)$
usually reaches its maximum at some $x^*$ between 0 and 1.
In the economics literature it is referred to
as Kelly-optimal investment ratio \cite{kelly1956}. It describes the optimal fraction of capital
that a prudent long-term investor should keep in relatively safe financial assets such 
as bonds or bank deposits while investing the rest in more risky assets such as stocks \cite{maslov1998}.
In our biological interpretation $x^*$ corresponds to the optimal fraction of the phage
population in the lysogenic state. At the Kelly-optimum the derivative of
$\Lambda$ with respect to $x$ is equal to zero, which is  realized at
\begin{equation}
\label{eq2}
x^{*}=p \cdot \frac{\Omega }{\Omega -1} 
\end{equation}
This equation is nearly identical to the Eq. 
8 in \cite{cohen1966} which 
expresses the optimal germination frequency of plant seeds
in environments where bad years would eliminate all germinated seeds. 
Furthermore, the positive correlation between $x^*$ and the frequency of 
environmental fluctuations $p$ predicted by the Eq. \ref{eq2} was 
confirmed by the empirical data on seed germination \cite{cohen1966,mayer}.

Before we proceed with analysis and modifications of our model we would like 
to highlight the main approximations/simplifications used throughout this study. 
Our first approximation is the use of the
Kelly theory, which implicitly assumes stochastic exponential growth. 
Thus in our simplified two-state model we ignore the (very real) dynamical
feedback between populations of phages and their bacterial hosts as the former
approach and ultimately reach steady state equilibrium. Such feedback in a 
system consisting of two types of phages (temperate and virulent) and one type of 
bacterial host (with susceptible, lysogenic, and resistant subpopulations) 
subject to variable nutrients
was analyzed in a classic paper  \cite{stewart1984}.
Stewart and Levin also compared purely virulent and 
temperate phage strategies in the steady-state of this closed system 
and reached general conclusions, which are in qualitative agreement with our predictions:
``Lysogeny is an adaptation for phage to maintain their populations in 
``hard times'', when the host bacterial density oscillates below that necessary 
for phage to be maintained by lytic infection alone.'' Here we confirm this
prediction using a very different approach, and expand it 
by calculating the optimal lysogenic fraction of a temperate phage.
Later on we will generalize our simplified
two-state model characterized by the generic  ``good'' or ``bad'' 
conditions to a multi-state model in which the current growth rate 
$\Omega(t)$ is drawn from an arbitrary (even continuous) probability
distribution. This variant of the model in principle allows 
us to consider the slowdown or even reversal in the growth rate of 
the phage population following the dynamical trajectory
derived in Ref. \cite{stewart1984}. Indeed, in our model we can 
represent this trajectory by the corresponding 
distribution of instantaneous growth rates $\Omega(t)$.
Abrupt downturns (``bad times'') in our model are 
caused by external events, such as e.g. invasions of 
new predators or emergence of resistant strains,  
(see e.g. Ref. \cite{thingstad1997}). In addition, the 
negative feedback between populations of the phage and its bacterial host 
\cite{stewart1984} can cause dramatic short term changes in 
the lytic growth rate.  
The last approximation behind the Eq. 1 is that
``bad-times" are treated as singular events of undetermined
duration, thereby ignoring the topic of the optimal rate 
for lysogen induction. We employ this simplification 
because the lysogenic state represents
a long term commitment that can be broken only due to 
rare stochastic fluctuations or excessive DNA damage 
of the host \cite{roberts1975}. 
This simplification is well justified for $\Omega \gg 1$ when 
the vast majority of phages were created during the 
previous time-step and thus the lysogenization frequency during this step 
directly determines the lysogenic fraction $x$ of the entire population. 
Therefore, in this study we focus exclusively on the choice 
between lytic and lysogenic states during the 
infection, and not on the small spontaneous release
of lytic phages from lysogens (of order $10^{-5}$ per generation
per bacteria for phage $\lambda$ \cite{baek}).
A more extended and general formalism allowing for discussion 
of both entry and exit rates in two-state phenotypic model 
can be found in  \cite{bergstrom2004,kussell2005,leibler2005,caswell2001}.
 
Phages are known to combine stochastic and regulated strategies 
for entry to lysogeny in response to a variety of external and 
internal signals \cite{lieb1953}. 
One example of such strategic response is provided by 
an increase in lysogenization frequency 
\cite{kourilsky1973} 
in response to reduced burst size 
when phages infect bacteria in a starved or stationary state 
\cite{sillankorva2004,weitz2008,wang2010}.
Here the short term optimization criterion 
would predict full virulence as long as $(1-p) \cdot \Omega>1$,
whereas the long term optimization (Eq. 2) 
would suggest a gradual increase
of the optimal lysogeny frequency $x^*$ 
as $\Omega$ is reduced and/or $p$ increases. 
In this study we do not consider the question 
of how phages can keep the lysogenic fraction 
of their population as close as possible
to its Kelly-optimal value $x^*$.  
Instead we concentrate on how $x^{*}$ itself depends on
environmental and biophysical parameters.

\subsection*{Regions of optimal temperate, virulent, or dormant phage strategies}

To simplify our calculations, above we assumed 
that during bad times the entire lytic phage population 
dies off and that the lysogenic subpopulation does not change at all.
Both assumptions can be relaxed by assuming a small but finite multiplicative ratio $\omega \ll 1$ 
quantifying the collapse (but not complete extinction) of the lytic subpopulation during bad times. 
We also introduce
the new parameter $\lambda$ for the growth 
(or decline) rate of the lysogenic subpopulation defined  
by the replication rate of their bacterial hosts. The mathematical 
approach developed in our study requires $\lambda$
to stay constant during both good and bad times 
in contrast to dramatic changes in growth rates of 
the lytic subpopulation.

In this more general case of our two-state model the 
logarithmic growth rate of the entire phage population is given by
$\Lambda(x)=(1-p) \log[(1-x) \Omega +x\lambda ]+p \cdot \log[(1-x)\omega + x\lambda]$ 
and the Kelly-optimal lysogenic fraction is given by (see Appendix for derivation).the 
\begin{equation}
x^* \; =\; p \cdot \frac{\Omega }{\Omega -\lambda} - (1-p) \cdot \frac{\omega}{\lambda-\omega} \label{eq3}
\end{equation}

The important new result is the existence of a finite threshold for 
transition between purely lytic (virulent) and mixed lytic-lysogenic 
(temperate) strategies of phages.  Assuming (quite realistically) that
$p \ll 1$, $\Omega \gg \lambda$, and $\omega \ll \lambda$ one gets 
the approximative relation 
$x^*=p-\omega/\lambda$, 
which predicts
that a
purely lytic strategy with $x^*=0$ is optimal when 
\begin{equation}
p<\frac{\omega}{\lambda} .
\end{equation}
In other words, virulent phages thrive when the probability of environmental 
downturns ($p$) is smaller than the relative impact of such downturns on 
lytic and lysogenic subpopulations ($\omega/\lambda$). On the opposite end of the spectra
the growth advantage of the lytic over the lysogenic state under good conditions shrinks as
$\Omega$ is decreased until it becomes comparable to $\lambda$.
In this case phages (as smart investors) should allocate 
progressively larger portion of their
population ``capital'' to the safety of the lysogenic state. 
When the time-averaged growth rate of a purely lytic 
population is less or equal than that of a purely lysogenic one, 
$(1-p) \Omega + p \omega \leq \lambda$, 
the Kelly-optimal lysogenic fraction is equal to 1 
instructing phages to permanently abandon the lytic strategy e.g. 
by transferring their genomes to plasmids.
In between these two extremes, for moderate likelihoods of 
bad times $p$, and substantial good times lytic growth rates 
$\Omega$, the temperate strategy will win. 
Thus the ``well-temperate phage" from our title is the one whose 
lysogenization frequency within the duration of a lytic burst cycle 
is approximately equal to the likelihood of the lytic population collapse:
$x^* \simeq p$.

\begin{figure}[h]
\includegraphics[trim = 0mm 0mm 0mm 0mm, clip,width=0.7\columnwidth]{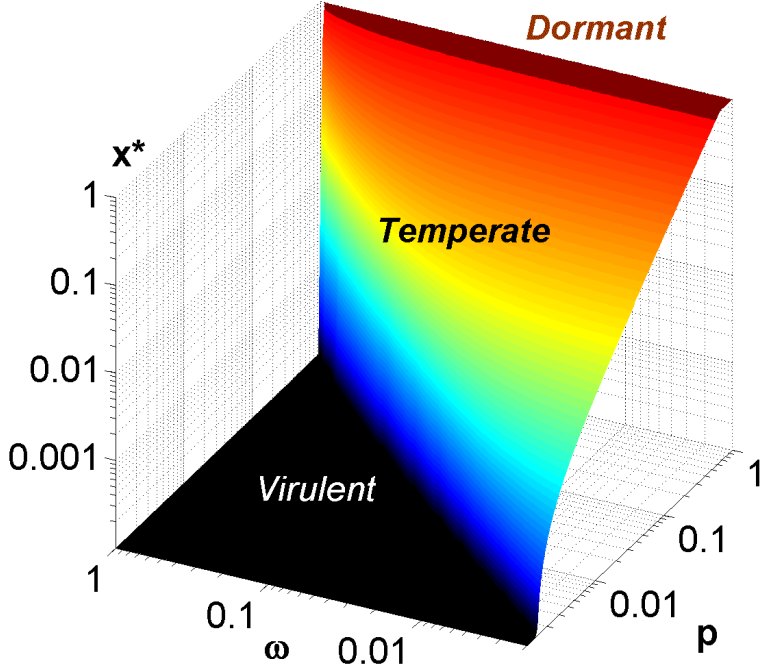}
\caption{\small\sl Kelly-optimal lysogenic ratio as a function of
$p$ - the probability of environmental downturn, 
and $\omega$ - the severity of population
collapse during such downturn. Note the sharp boundary separating
purely virulent (the blue region in the upper left corner)
and temperate strategies.
Equally sharp boundary separating temperate ($0<x^{*}<1$) and
dormant ($x^*=1$) strategies is less visible in this plot because of selection 
of colors.   Here we used a two-state model with $\Omega=3$ and
$\lambda=1$ but other values of these
parameters do not change the qualitative
picture shown here.
}
\label{figure2}
\end{figure}

The plot of the Kelly-optimal lysogenic fraction $x^{*}$ as a function of the 
probability - $p$ and the severity - $\omega$ of population collapses during bad times
is shown in Fig. \ref{figure2}.

One can further generalize our model from two-state environments 
to multiple or even continuous state environments.  In this model the lytic population 
growth rate $\Omega(t)$ during a given time-interval $t$ is drawn from an 
arbitrary probability distribution $\pi(\Omega(t))$. The two-state model considered above
corresponds to $\pi(\Omega(t))=(1-p)\delta(\Omega(t)-\Omega)+p\delta(\Omega(t)-\omega)$.

In the appendix we prove the existence of a unique 
Kelly-optimal strategy which is
\begin{eqnarray}
\qquad \mathrm{Temperate \; (0<x^{*}<1) \; }&\mathrm{for}&  \;\;\langle \Omega \rangle > \lambda  \;\;\; \mathrm{and} \;\;\;
\langle \frac{1}{\Omega} \rangle > \frac{1}{\lambda} 
\label{temperate_conditions} \\
\mathrm{Virulent \quad (x^{*}=0) \quad }&\mathrm{for}& \;\quad \qquad \langle \frac{1}{\Omega} \rangle \le  \frac{1}{\lambda} \qquad  \label{virulent_conditions} \\
\mathrm{Dormant \quad (x^{*}=1) \quad }&\mathrm{for}& \; \mathrm\quad \qquad \langle \Omega \rangle \le \lambda  \qquad .
\label{dormant_conditions}
\end{eqnarray}

Here $\langle \rangle$ 
denotes the long-term time average calculated using $\pi(\Omega(t))$.
When $\langle \Omega \rangle \le \lambda$ (Eq. \ref{dormant_conditions})
there is no growth advantage
of being lytic and hence the optimal 
choice is for the entire phage population to go dormant 
into the lysogenic state: $x^{*}=1$.
On the other hand, when 
$\langle 1/\Omega \rangle  \le 1/\lambda$ (Eq. \ref{virulent_conditions}), 
the frequency and severity of  lytic population collapses is not sufficient to justify even a marginal 
investment into "safe" lysogenic state. Hence the optimal choice in this limit 
is for the entire phage population to remain lytic: $x^*=0$. 
In between these two extreme scenarios the temperate strategy is optimal 
(Eq. \ref{temperate_conditions}), and, as shown in the Appendix, 
the ``well-temperate'' (Kelly-optimal) lysogenic ratio $x^*$
is determined by numerical solution to 
\begin{equation}
\langle \frac{1}{(1-x^*)\Omega+x^*\lambda }\rangle 
=\frac{1}{\lambda} \qquad .
\label{eq6}
\end{equation}

If the lysogenic subpopulation was completely stable,  
one would have the following paradox:
the growth in the Kelly-optimal solution can only be larger
or equal than the lysogenic growth rate $\lambda$. Thus, for 
$\lambda=1$, the optimal temperate strategy would necessarily be non-stationary.  
This paradox does not exist if $\lambda<1$. Such smaller growth of lysogenized hosts 
is in fact expected due to either cost of integrated prophages (see Ref. \cite{stewart1984} for 
quantitative estimates)  or predation of lysogens by other types of phages.
In this case we expect the entire phage population to self-organize 
into a stationary state where over the longest times scale it 
neither shrinks nor grows:
$\langle \log [(1-x^*) \Omega +x^*\lambda] \rangle =0$. 
%
%

\subsection*{Independent, interconnected environments favor virulence}

Besides previously discussed factors such as the frequency and the severity of environmental collapses,  
the choice between virulent and temperate strategies depends also on phages' ability to access 
(e.g. by diffusion) multiple spatially-separated environments, which are 
fluctuating independently of each other.
  
In general, the access to multiple independently fluctuating environments favors the virulent strategy, 
in much the same was as the access to a well-diversified  
investment portfolio consisting of multiple independently fluctuating stocks reduces investor's 
risk exposure  \cite{marsili1998} and tempts to move his/her capital out of the safety of a 
low interest bank deposit. Analogous trends have been reported  \cite{bulmer1984} for seed 
germination of plants, also suggesting that the fast growing but risky strategy wins when offsprings 
can be spread between many independently fluctuating environments.

For phage populations, as their access to multiple independent environments increases, hedging of bets by 
the lysogeny becomes progressively less and less important. Eventually, in the limit of a large number of 
strongly  interconnected environments one expects 
the purely lytic (virulent) strategy to win over any temperate strategy. 

To quantify this common sense prediction in terms of phage biophysics and environmental 
parameters, we mathematically consider the case where the diffusion connects
phage populations in $N$ of independent yet statistically identical
environments. We assume that at every
time-step the diffusion distributes a fraction $\gamma < 1$ of the entire phage population
equally across all environments.  

Generally speaking, in the multi-environment model 
the diffusion constant $\gamma$ takes the role of the collapse 
ratio $\omega$ in the single-environment model.
Indeed, during severe environmental downturns when the entire lytic subpopulation 
is lost it is replenished by phages diffusing from other environments at the rate $\gamma$. Therefore, in our subsequent numerical simulations we set $\omega=0$. 

The multiplicative growth ratio of the entire {\it lytic} 
subpopulation in all environments is given by 
$\Omega_{total}(t) = \sum_{i=1}^N \Omega_i(t) P_i(t)/\sum_{i=1}^N P_i(t)$.
where $P_i(t)$ is the accumulated lytic subpopulation in the environment $i$ and
$\Omega_i(t)$ is its growth ratio at time-step $t$. 
As before, it is equal to $\Omega$ with probability $1-p$ and $0$ with probability $p$. 
To estimate the overall growth, one needs to know the distribution
of populations in all environments, which in turn is dependent
on the spatiotemporal pattern of growths and collapses.
Assuming the same lysogenic ratio $x$ and lysogenic growth rate $\lambda$
in all environments , the long-term logarithmic growth rate of the total phage population 
(both lytic and lysogenic) 
is given by the time average 
$\langle \log((1-x)\Omega_{total}(t) +x\lambda) \rangle$.

Numerical simulations allowed us to estimate the distribution of $\Omega(t)$
and to subsequently numerically  solve the Eq. \ref{eq6} 
(see "Model and Numerical Simulations" for more details)  to self-consistently 
determine Kelly-optimal lysogenic fraction $x^*$. 
Fig. 3A show how thus defined $x^*$ depends on the number of environments 
$N$ and the diffusion constant $\gamma$ in a model with $\Omega=3$ and $p=0.1$. 
This figure quantifies the common-sense 
prediction that $x^*$ should decrease
with both $N$ and $\gamma$. The same overall trend
was previously found in a similar model \cite{bulmer1984}
for plant seed germination and dispersal. 
In Fig. 3B we show that in the model with $\gamma=10^{-4}$ and 
$N=100$ when the lytic growth ratio $\Omega$ goes up 
and/or the collapse frequency $p$ goes down 
the optimal lysogeny frequency $x^*$ decreases and ultimately 
vanishes, indicating a transition to the purely virulent strategy. 
In contrast to a rather weak dependence
of $x^*$ on $\Omega$ predicted for a single isolated environment (see Eq. \ref{eq3}), 
our numerical results for multiple interconnected environments shown in Fig. 3B 
indicate that the fast lytic growth quantified by $\Omega$ strongly favors purely virulent strategy. 
Our simulations demonstrate the existence of a sharp boundary 
(see the border between colored and black areas in both panels of Fig.3)
above which the lysogeny is no longer required and 
pure virulence becomes the optimal phage strategy.
\begin{figure}[H]
\includegraphics[trim = 0mm 0mm 0mm 0mm, clip,width=0.5\columnwidth]{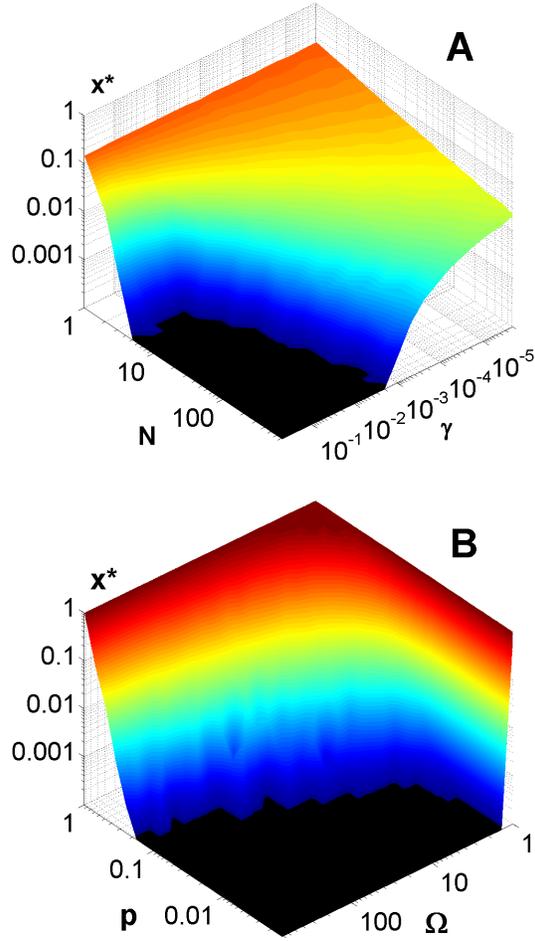}
\caption{\small\sl Kelly-optimal lysogenic ratio $x^{*}$ plotted 
as a function of the diffusion rate $\gamma$ and the number of environments $N$ 
for a two-state model with $\Omega=3$ and $p=0.1$ (panel A) 
or as a function of the lytic growth ratio $\Omega$ and the 
probability of population collapse $p$ for model with 
$\gamma=10^{-4}$ and $N=100$.
Note the sharp boundary to purely virulent strategy with $x^*=0$ (black area) 
which is optimal for large 
$\gamma$ and $N$, as well as for large $\Omega$ and small $p$.
The temperate strategy (colored area) is optimal in the opposite limit.}
\label{figure3}
\end{figure}

\section*{Discussion}
Sustainability of different strategies of phage predation has been 
theoretically explored before. Particular attention has been paid
to the question of  long-term sustainability of the pure virulent strategy
\cite{campbell1960,levin1977,weitz2008,wang2010,heilman2012}.
At a first glance populations of lytic phages are inherently prone to collapses. 
Indeed, each infection leads to hundreds of new
phage particles, which rapidly deplete the population of 
susceptible bacteria leading to the steady state with 
a low density of hosts prone to collapses \cite{stewart1984}. 
Temperate phages, on the other hand, 
provide lysogens with the immunity against their own siblings and therefore are able 
to survive irrespective of host's density. 
Models of interactions between temperate phages and their hosts 
have been considered by   \cite{noack1968,paynter1970,wang2010}.
Stewart and Levin \cite{stewart1984} directly compared 
the resilience of temperate and virulent phage populations after 
their hosts were exposed to changes in nutrient levels. One of the 
important results of that study is that for substantial variations of 
nutrient availability temperate phages fare better than virulent ones.
We have confirmed this earlier prediction using a very different 
type of model that takes into account not just nutrient variability but any 
other type of environmental downturns. Whenever temperate strategy 
beats virulent and dormant ones our approach allowed us to show that the 
optimal lysogeny frequency of a  ``well-temperate'' phage resulting in the fastest 
population growth, has to be close to the probability for collapse of their 
local environments.

For pedagogical reasons, in our presentation above we assumed that the 
lysogenization frequency $x$ has to stay the same throughout the duration 
of good environmental conditions. 
In other words, we ignored phages ability to 
actively sense the environmental state and 
to use this information to dynamically 
adjust the lysogenization frequency.
However, our estimates can be directly extended to the case where phages 
can distinguish between multiple types of ``good'' environments each characterized 
by its growth rate and its own likelihood and intensity of collapse.
If these environments occur randomly and independently from each other, 
the overall growth can be factorized and
we predict phages would adjust the lysogenization frequency in each one of these 
environments to be given by Eq. \ref{eq3} with parameters $\Omega$, $p$, and $\omega$ 
characteristic of this particular environment.
One example of this is provided  by $\lambda$-phages collecting 
information about the nutritional state of the host 
and on whether it was 
simultaneously co-infected by other 
$\lambda$-phages \cite{kourilsky1973}. 
This information is then processed by the phage to make 
lysis-vs-lysogeny decision that is in agreement with 
our predictions. 
That is to say, $\lambda$-phage's lysogenization frequency is known 
to increase \cite{kourilsky1973} when they infect starved or multiple-infected hosts 
which both signal reduced prospects of lytic growth captured in our 
model by reduction in $\Omega$ and/or increase in $p$.

Optimal behavior of phages with respect to choice between  
virulent  and temperate strategies depends on multiple
extrinsic and intrinsic parameters such as 
phages' burst size, host range, hosts' availability, susceptibility, and average growth rate, 
the frequency and severity of environmental collapses, and finally phages' ability to 
diffuse across multiple environments within their lifetime as an infectious particle. Our key predictions are: 
1) The temperate phage strategy dominates when environmental downturns 
happen  frequently, are severe, or when phages live in isolated and simple environments. 
2) The virulent phage strategy gains the upper hand when the probability of 
downturns gets smaller, and/or collapses themselves are milder, 
and phages have access to multiple hosts or environments connected 
by diffusion (see Fig. 3B). This prediction is consistent with the empirical 
observation \cite{paepe2006} 
that virulent phages have systematically larger adsorption rates 
and shorter latency times than temperate phages. 

Finally, our analysis suggests a simple explanation of why virulent mutants of
temperate phages are not commonly found in the wild. 
Assuming that mutants are exposed to the same environmental risks
as their ancestors we can calculate the reduction in the growth rate 
of a virulent mutant relative to its well-tempered ancestor:
\begin{equation}
\Delta \Lambda(\mathrm{virulent \, mutant})
\simeq p \log(\frac{\omega}{p} )
\end{equation}
(see SI for the exact formula and its derivation). 
This is a particular case of Bergstrom and Lachmann's results 
\cite{bergstrom2004}
relating fitness differences to Shannon entropy of the environment and 
the measure of organism's information about the environment. 
In our case fitness loss of the virulent mutant compared to its 
optimally-tempered wild type ancestor is related to the information lost 
when the probability of environmental collapse - $p$ is used 
as a proxy for its severity -  $\omega$.

The temperate strategy is optimal when $p>\omega$ (see Eq. \ref{eq3}).
Hence, a virulent mutant of a temperate phage has lower fitness than its ancestor: 
$\Delta \Lambda(\mathrm{virulent \, mutant}) <0$. 
Overcoming this fitness barrier by adjusting any of the 
other intrinsic properties of the phage such as e.g. its host range would 
require multiple simultaneous mutations and is, therefore, unlikely.
This prediction of non-sustainability of virulent mutants of
temperate phages is in agreement with the 
observation that protein families in known virulent phages have 
essentially no functional overlap with those in temperate phages
\cite{lima-mendez2008}.

{\color{black}
The bet-hedging approach introduced in our study 
focuses on random external shocks to the system
at the expense of its intrinsic dynamics. 
Thus we treat the growth rates in a given environment as 
fixed and completely ignore the density-dependent 
feedback between populations of phages and their bacterial hosts
\cite{campbell1960,levin1977,stewart1984}.
Neglecting such feedback allowed us to obtain multiple 
mathematical insights that would be difficult to derive otherwise.
However, this simplification may 
influence some of our predictions, in particular 
for ``smart'' phages capable of using the state of its host 
to predict availability of hosts in near future.
Future work is needed to combine our bet-hedging 
formalism with density-dependent population dynamics.
}

\section*{Methods}
\subsection*{Model and numerical simulations}
Our model is updated in discrete time-steps. At each time-step the lytic subpopulation in each of the environments 
is either grows by a factor $\Omega>>1$ (with probability $1-p$),
or collapses by a factor $\omega<<1$ (with probability $p$). 
The lysogenic subpopulation is initially kept constant, and subsequently re-adjusted such that
the selected fraction $x$ of the total phage population in a given environment 
is assigned to the lysogenic state.  
In case of multiple independent environments (Fig. 3), 
the diffusion operates at each times-step by 
redistributing the fraction $\gamma$ of the total phage population 
equally among all environments: $P_i (t) \to P_i(t)(1-\gamma)+\gamma \sum_{j=1}^N P_j(t)/N$.
The simulations provide us with the numerical expression for the 
distribution of populations $P_i$  across the environments and times.  This distribution 
in its term determines the distribution 
$\pi(\Omega_{total}(t))$ of growth rates of the global 
phage population in all of the environments.  The Kelly-optimal fraction $x^*$ 
is then found by numerically solving the Eq. \ref{eq6} for the distribution 
$\pi(\Omega_{total}(t))$.  After this we recalculate 
the distribution of $P_i$ and $\pi(\Omega_{total}(t))$ for the new value of $x=x^*$.
This iterative process is repeated until the relative difference of $x^*$ 
during subsequent iterations is less than 1\%.
\subsection*{Kelly-optimal ratio for the two-state environmental model}

In the general version of the two-state environmental model the long-term
logarithm growth rate is given by
$\Lambda(x)=(1-p) \cdot \log((1-x)\Omega+x\lambda)+p \cdot \log((1-x)\omega+x\lambda)$.
Here $\Omega$ and $\omega$ are the growth rates of the lytic subpopulation under
good and bad environmental conditions correspondingly, while $\lambda$ is the constant growth rate
of the lysogenic subpopulation. 
Taking the derivative with respect to $x$ and setting it to $0$ results in the following
equation for $x^*$
$$0=\frac{(1-p)(\lambda-\Omega)}{\Omega (1-x^*)+\lambda x^*}+
\frac{p(\lambda-\omega)}{\omega (1-x^*)+ \lambda x^*} \qquad , $$ 
which can be further simplified to
$$0=\frac{1-p}{x^*-\Omega/(\Omega-\lambda)}+
\frac{p}{\omega/(\lambda-\omega) +x^*} \qquad .$$
Grouping all the terms with $x^*$ on one side results in the following
expression for the Kelly-optimal lysogenic fraction:
\begin{equation}
\label{supp_eq3}
x^* \; =\; p \frac{\Omega }{\Omega -\lambda} - (1-p) \cdot
\frac{\omega}{\lambda-\omega} \quad .
\nonumber
\end{equation}

\subsection*{Kelly-optimal ratio for the multi-state (continuous) environmental model}

Here we consider a more general model in which the current growth rate $\Omega(t)$ 
of lytic subpopulation is not limited to just 
two values $\Omega$ and $\omega$ but is independently drawn from an arbitrary probability 
distribution $\pi(\Omega(t))$. In this case the long-term logarithmic growth rate  is given by
\begin{equation}
\Lambda(x)=\int \log((1-x)\Omega+x\lambda) \pi(\Omega) d\Omega \qquad .
\end{equation}
The Kelly-optimal lysogenic ratio $x^{*}$ is determined by solving
\begin{equation}
0=\frac{d \Lambda}{dx}=\int \frac{\lambda-\Omega}{(1-x^{*})\Omega +x^{*}\lambda } \pi(\Omega) d\Omega
\label{eq_x_cont} \qquad .
\end{equation}
Note that the second derivative of $\Lambda(x)$ equal to
\begin{equation}
\frac{d^2 \Lambda}{dx^2}=-\int \frac{(\Omega-\lambda)^2}{((1-x)\Omega+x\lambda)^2} \pi(\Omega)
d\Omega
\end{equation}
is always negative. The boundary conditions are given by
$d \Lambda/ dx |_{x=0}=\langle (\lambda-\Omega)/\Omega \rangle $ $=\lambda \langle 1/\Omega \rangle-1$
and $d \Lambda/dx |_{x=1}=1-\langle \Omega \rangle/\lambda$. Thus, as long as
$\lambda \langle 1/\Omega \rangle-1>0$ (or $\langle 1/\Omega \rangle > 1/\lambda$ )
and $1-\langle \Omega\rangle/\lambda<0$ (or $ \langle \Omega\rangle>\lambda$) 
a unique solution for the Kelly-optimal
lysogenic fraction $x^{*}$ between 0 and 1 exists.

When $\langle \Omega \rangle \le \lambda$
there is no growth advantage (yet all the risks) of going lytic and hence the optimal state
for the phage population is to be 100\% lysogenic: $x^{*}=1$.
On the other hand when $\langle 1/\Omega \rangle  \le 1/\lambda$ the growth rate
during bad times (small $\Omega$) dominating this average is not low enough to justify
even marginal ``safety net investment'' into the lysogenic state. Hence the
optimal strategy for phage population in this case is to be 100\% lytic: $x^{*}=0$.

A more concise way to write the equation for the Kelly-optimal lysogenic ratio can be
derived by multiplying the Eq. \ref{eq_x_cont} by $1-x^*$ and 
noticing that the numerator can be regrouped as
$(1-x^*) (\lambda-\Omega)=\lambda-[(1-x^*)\Omega+\lambda x^*]$.
Hence, the ratio under the integral of the Eq. \ref{eq_x_cont} can be
replaced with $\lambda/[(1-x^*)\Omega+x^*\lambda]-1$ which gives
\begin{equation}
\langle \frac{1}{(1-x^*)\Omega+x^*\lambda}\rangle =
\int \frac{1}{(1-x^*)\Omega+x^*\lambda} \pi(\Omega) d\Omega=\frac{1}{\lambda} \qquad .
\nonumber
\end{equation}
Hence the purpose of $x^*$
is to provide an ``insurance'' lower bound $x^*\lambda$ for the denominator 
when $\Omega$ is very small. 
This way the Kelly-hedged growth ratio $\Omega^*=(1-x^*)\Omega+x^*\lambda$
satisfies
$\langle \frac{1}{\Omega^*} \rangle = \frac{1}{\lambda}$.
As derived above such insurance is necessary only when 
small values of $\Omega$ happen sufficiently frequently to make
$\langle \frac{1}{\Omega} \rangle > \frac{1}{\lambda}$.

For log-normally distributed $\Omega$ described by
$\pi (\Omega)=\exp(-(\log \Omega -\mu)^2/2\sigma^2)/\Omega/ \mathrm{Norm}$ 
the calculation of the parameter range 
for which temperate strategy is Kelly-optimal is especially simple.
Indeed in this case $m$-th moment of $\Omega$, $\langle \Omega^m \rangle=
\exp(m \mu +m^2 \sigma^2/2)$.
Thus, in order to have $\langle \Omega \rangle > \lambda $ and $\langle \Omega ^{-1} \rangle > 1/\lambda$ 
one needs  $\log \lambda -\sigma^2/2 < \mu <\log \lambda+\sigma^2/2$. Adding $\sigma^2/2$ to all sides of this 
double inequality one gets a condition for optimality of the temperate strategy as
\begin{equation}
\log \lambda <\log \langle \Omega \rangle < \log \lambda +\mathrm{Var} 
(\log \Omega)=\log \lambda  +\sigma^2  \label{kelly_ineq_lognormal} \qquad .
\end{equation}
In other words, in order for the temperate strategy to beat the virulent and the dormant ones, 
the logarithm of the average growth rate in the lytic state 
has to be within one standard deviation of $\log \Omega$ above the 
logarithm of the growth rate of the lysogenic state.
This is possible either when these two growth rates are very close to each other or when the variability of $\log \Omega$ 
is very large. Note that this equation can be realized when $\lambda<1$, $\langle \Omega \rangle>1$, and the overall 
phage population is stationary: $(1-x^*)\langle \Omega \rangle+x^*\lambda =1$.

\subsection*{Fitness advantage of the Kelly-optimal strategy over purely 
virulent strategy in the two-state environmental model}

To quantify the fitness advantage of the Kelly-optimal strategy over purely 
virulent strategy in the most general formulation of the two-state environmental 
model let's recall that
\begin{equation}
\Lambda(x)=(1-p) \log [\Omega(1-x)+\lambda x]+p\log [\omega(1-x)+ \lambda x] \qquad .
\end{equation}
Thus the fitness advantage $s$ of the Kelly-optimal over purely virulent strategy
can be written as
\begin{eqnarray}
s &=&\Lambda(x^*)-\Lambda(0)=\nonumber \\
&=&(1-p) \log [1-x^*+\frac{\lambda}{\Omega} x^*]+p\log [1-x^*+ \frac{\lambda}{\omega} x^*]=\nonumber \\
&=&(1-p) \log [1-x^*\frac{\Omega-\lambda}{\Omega}]+p\log [1+x^* \frac{\lambda-\omega}{\omega}] \qquad.
\end{eqnarray}
Recalling, that according to Eq. \ref{supp_eq3}
$$
x^* \; =\; p \frac{\Omega }{\Omega -\lambda} - (1-p) \cdot
\frac{\omega}{\lambda-\omega} \qquad ,
$$
one can further simplify the expression for $s$ to be exactly equal to
\begin{equation}
\label{supp_eq2}
s = (1-p) \log[(1-p) \cdot (1+\tilde{\omega})]+p \log[p \cdot (1+1/\tilde{\omega})] \qquad ,
\end{equation}
where $\tilde{\omega}$ is a shorthand for
\begin{equation}
\tilde{\omega}=\frac{\omega(\Omega-\lambda)}{\Omega(\lambda-\omega)}
\end{equation}
Perhaps the most concise expression for $s$ is in terms of the Kullback-Leibler relative entropy
$S_{K-L}$ between bimodal probability distributions $p$ and $q$:
\begin{equation}
s = S_{K-L}= (1-p) \log \frac{1-p}{1-q} + p \log \frac{p}{q} \qquad ,
\end{equation}
where the ``probability'' $0<q<1$ is given by the ratio
\begin{eqnarray}
q&=&\frac{\tilde{\omega}}{1+\tilde{\omega}}=\nonumber \\
&=&\frac{\omega(\Omega-\lambda)}{\Omega(\lambda-\omega)+\omega(\Omega-\lambda)}=\nonumber \\
&=&\frac{\omega}{\omega+ (\lambda-\omega)\frac{\Omega}{\Omega-\lambda}}
\end{eqnarray}
In the limit $\Omega \gg \lambda$ one can further approximate
\begin{equation}
q=\frac{\omega}{\lambda} \qquad ,
\end{equation}

\subsection*{Acknowledgments}
Work at Brookhaven was supported by grants PM-031 from the Office of 
Biological Research of the U.S. Department of Energy. Work at Copenhagen was supported by the Danish National Research Foundation. 

\noindent
{\bf Author Contributions Statement:} SM and KS participated 
equally in all parts of this work.\\\\
{\bf Additional Information:} The authors declare no competing financial interest. 
\end{doublespace}
\end{document}